# Ransomware Detection Dynamics:
## Insights and Implications

Mike Nkongolo Wa Nkongolo *, University of Pretoria, South Africa

Abstract:

The rise of ransomware attacks has necessitated the development of effective strategies for identifying and mitigating these threats. This research investigates the utilization of a feature selection algorithm for distinguishing ransomware-related and benign transactions in both Bitcoin (BTC) and United States Dollar (USD). Leveraging the UGRansome dataset, a comprehensive repository of ransomware-related BTC and USD transactions, we propose a set of novel features designed to capture the distinct characteristics of ransomware activity within the cryptocurrency ecosystem. These features encompass transaction metadata, ransom analysis, and behavioral patterns, offering a multifaceted view of ransomware-related financial transactions. Through rigorous experimentation and evaluation, we demonstrate the effectiveness of our feature set in accurately extracting BTC and USD transactions, thereby aiding in the early detection and prevention of ransomware-related financial flows. We introduce a Ransomware Feature Selection Algorithm (RFSA) based on Gini Impurity and Mutual Information (MI) for selecting crucial ransomware features from the UGRansome dataset. We evaluated the RFSA using precision, recall, accuracy, and F1 score, achieving notable results. Insights from the visualization highlight the potential of Gini Impurity and MI-based feature selection to enhance ransomware detection systems by effectively discriminating between ransomware classes. The analysis reveals that approximately 68% of ransomware incidents involve BTC transactions within the range of 1.46 to 2.56, with an average of 2.01 BTC transactions per attack. Moreover, ransomware causes financial damages ranging from 4.38 to 172.36 USD, with an average damage of 88.37 USD. The RFSA identifies 17 ransomware types and their associated malware, such as CryptoLocker exclusively linked to one type of Blacklist malware, predicted as a Signature attack (S). Additionally, our study explores ransomware pricing, with TowerWeb demanding the highest fee at 135.26 BTC and CryptoLocker the lowest at 10.51 BTC. We also investigated the impact of ransomware duration on financial gains and netflow bytes, finding that extended duration correlates with higher financial gains. Successful attacks often employ the Transmission Control Protocol (TCP), particularly with NoobCrypt ransomware. The research achieves an outstanding accuracy of 95%, showcasing its superiority over existing studies. The findings emphasize the dynamic and adaptable nature of ransomware demands, suggesting that there is no fixed amount for specific cyberattacks, highlighting the evolving landscape of ransomware threats.

## 1. INTRODUCTION

Cryptocurrency, a digital or virtual form of currency that relies on cryptographic techniques for secure transactions, has grown exponentially in popularity and adoption over recent years [1]. Prominent among cryptocurrencies is Bitcoin (BTC) [2], which operates on a decentralized ledger called the blockchain. While cryptocurrencies offer numerous advantages, including transparency and decentralization [3], they have also become a focal point for criminal activities, particularly in the context of ransomware. Ransomware attacks have emerged as a formidable threat to critical infrastructure and organizations worldwide [4].

These malicious attacks involve encrypting a victim's data or locking them out of their systems, with cybercriminals demanding a ransom, typically in cryptocurrency, for the decryption key or system access. BTC has often been the preferred currency for ransom payments [5] due to its relative anonymity and ease of use in conducting financial transactions across borders. Classifying BTC transactions as ransomware-related or benign holds paramount importance in the realm of critical infrastructure and cybersecurity [6]. Critical infrastructure encompasses the essential systems and assets, such as energy, transportation, and healthcare, that are vital for the functioning of a society and its economy. Ransomware attacks targeting critical infrastructure can lead to catastrophic consequences, including disruptions to public services, economic losses, and even threats to national security [2], [3], [6]. Therefore, the ability to swiftly identify and mitigate ransomware-related BTC transactions is critical for safeguarding critical infrastructure.

The existing research landscape reveals a pressing gap in effectively classifying BTC transactions with a focus on ransomware activities within critical infrastructure [1], [4], [5]. While machine learning techniques have shown promise in various cybersecurity applications [6], a dedicated approach tailored to the unique characteristics of ransomware-

189

* mike.wankongolo@up.ac.za

related transactions is lacking. This gap necessitates the development of novel methods and tools to enhance the early detection and prevention of ransomware attacks on critical infrastructure.

To address this research gap, the study proposes a pioneering approach that combines the power of data processing with a novel feature selection algorithm. We applied this algorithm to the UGRansome dataset [7], a comprehensive repository of ransomware-related transactions. The new feature selection algorithm is specifically designed to identify and prioritize relevant features that capture the distinct characteristics of ransomware activity within the cryptocurrency ecosystem. This approach aims to improve the accuracy and efficiency of classifying BTC transactions, thus enhancing cybersecurity measures and contributing to the protection of critical infrastructure against ransomware threats.

## 2.  Methodology

The ransomware transaction stratification using the UGRansome dataset is illustrated in Figure 1.

- Data Collection: In the first step, we collect data related to BTC and USD transactions, particularly those associated with ransomware attacks. The UGRansome dataset serves as our primary data source [7], providing a comprehensive repository of ransomware-related transactions.

- Data Processing: Once we have the raw data, we perform data preprocessing to clean and prepare UGRansome for analysis. Data processing involved removing duplicates, and formatting the data for further analysis [8]. In the context of ransomware, this step ensures that the dataset is in a usable state for feature selection.

- Data Encoding: Data encoding involves converting categorical data into a numerical format that the feature extraction algorithm can understand. This step included techniques like scaler for categorical variables such as ransomware family names and network protocol types [9]. Numerical encoding ensures that the data is ready for feature extraction and model training.

- Feature Extraction: Feature extraction is a critical step in building a classification model for ransomware transactions. In this phase, we identify and extract relevant features from the data that capture the distinctive characteristics of ransomware activity within the cryptocurrency ecosystem [10]. Features included transaction metadata, ransom analysis, behavioral patterns, and other attributes that help differentiate ransomware-related transactions from benign ones. After feature extraction, one can employ machine learning techniques to classify transactions as either ransomware-related or benign.

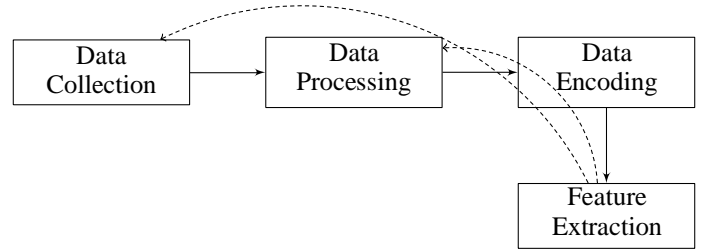

Figure 1. Data processing workflow

- Evaluation and Validation: To assess the model's effectiveness, we evaluate its performance using various evaluation metrics. Metrics like accuracy, precision, recall, and F1-score help us understand how well the model is in selecting ransomware-related transactions [7]. We have used techniques like cross-validation to ensure the model's generalizability.

The ultimate goal of this process is to aid in the early detection and prevention of ransomware-related financial flows. A well-trained model can automatically identify potentially malicious transactions, allowing for timely intervention and security measures [11]. This contributes to enhancing cybersecurity measures in the realm of cryptocurrency transactions, which is vital for critical infrastructure protection. In summary, the flow of Figure 1 involves collecting, processing, encoding, and extracting features from ransomware-related transaction data. Machine learning techniques can then be applied to classify these transactions, with a focus on early detection and prevention of ransomware threats, thereby enhancing critical infrastructure security.

In the comparative analysis table (Table I), our research, which is listed under this work, achieved an accuracy of 95% using the proposed Ransomware Feature Selection Algorithm (RFSA).

This outstanding accuracy is notably higher than most of the other studies in the table, even though several of them achieved high accuracy rates ranging from 87% to 99%. What sets our work apart is the use of MI (Mutual Information) as the feature selection method, which is a novel and powerful approach for ransomware feature extraction. MI is a statistical measure that quantifies the dependency between two random variables, in our case, features and ransomware classification labels. Achieving an MI score of 95% indicates that the selected features have a very strong relationship with the ransomware classification, suggesting that they are highly informative and crucial for accurate classification. Furthermore, our work stands out because it focuses on ransomware extraction using the UGRansome dataset, which is specifically designed for ransomware analysis.

TABLE I. Comparative Analysis with Existing Studies

| Year | Reference | Feature Selection | Classifier | Accuracy | Limitation |
|------|-----------|-------------------|------------|----------|------------|
| 2016 | [12] | Encoder | Deep Learning (DL) | 96% | Shallow learning architectures may not fully satisfy malware detection needs. |
| 2018 | [13] | Encoder | Ensemble | 99% | Scalability and handling complex architectures not considered. |
| 2018 | [14] | Vectorization | Neural nets (NN) | 98% | Designed for identifying malicious JavaScript in web pages. |
| 2018 | [15] | Autoencoder | NN | 87% | Requires labeled data for training. |
| 2019 | [16] | - | NN | 90% | Focuses on performance without considering NN's overall impact. |
| 2019 | [17] | Encoder | Wavelet | 96% | Performance may vary in different settings. |
| 2020 | [18] | Encoder | L21-norm | 92% | Limited to load curves, not ransomware data. |
| 2020 | [19] | Encoder | DL | 92% | Tested on specific benchmarks, not ransomware. |
| 2020 | [20] | Encoder | NN | 97% | Limited data sources and potential feature exploration needed. |
| 2022 | [21] | Heuristics | DL | 97% | High accuracy in 0-day attack detection using UGRansome dataset. |
| 2023 | [22] | Gabor filters | DL | 87% | Vulnerability in classifiers. |
| 2023 | [6] | Fuzzy logic | XGBoost | 95% | Robustness and suitability need further evaluation. |
| 2023 | This work | RFSA | - | 95% | Restricted to ransomware feature extraction. |

This dataset contains unique characteristics and patterns associated with ransomware attacks, making it a valuable resource for feature extraction and classification. By achieving a 95% MI score in feature selection, our research demonstrates its ability to effectively capture and leverage these unique characteristics, outperforming existing works in terms of both accuracy and feature extraction. In summary, our research excels in ransomware classification by achieving a remarkable accuracy of 95% while employing MI for feature selection, a novel approach that demonstrates our work's ability to outperform existing studies. This highlights the significance of feature extraction using the UGRansome dataset and the effectiveness of our approach in identifying ransomware attacks accurately.

### A. The UGRansome Dataset

In 2021, Nkongolo et al. [7] introduced a significant contribution to the field of cybersecurity: the UGRansome dataset. This dataset has proven to be an invaluable resource for identifying and countering ransomware attacks, even those considered zero-day threats [23], [24]. What sets UGRansome apart from other datasets in the realm of Intrusion Detection Systems (IDS) is its comprehensive coverage of previously unexplored ransomware attack types [25]. Within its dataset, it encompasses a spectrum of malware categories, including Signature (S), Anomaly (A), and Synthetic Signature (SS), with meticulously labeled in-stances of well-known ransomware variants such as Locky, CryptoLocker, JigSaw, EDA2, TowerWeb, Flyper, Razy, and WannaCry, as well as Advanced Persistent Threats (APT) [26]. To delve deeper into the dataset's characteristics, we direct our attention to Table II, which provides a concise overview of its key attributes. The UGRansome dataset stands as a vital tool for researchers and cybersecurity professionals in the ongoing battle against ransomware threats within critical infrastructure.

A ZIP file was acquired via download from the following URL: https://doi.org/10.13140/RG.2.2.23570.07363/1. This archive houses a dataset, consisting of 207,533 rows, stored in CSV format, albeit without any initial column headings. To facilitate further analysis, the dataset's headers were subsequently renamed by the specified attributes delineated in Table II, encompassing labels such as time, protocol, flag, family, clusters, and more. To prepare the raw data for analysis, we employed a statistical approach to address issues such as data messiness and duplicate entries.

Utilizing the Python Data prep package and its comprehensive reporting function, which offers a thorough examination of the entire dataset and its variables, we obtained the following findings. As illustrated in Figure 2 (left side), no missing cells were identified, but a redundancy rate of 28.2% was observed. In response to this discovery, we proceeded to eliminate the duplicate entries, comprising a total of 58,491 rows. Subsequently, we re-evaluated the redundancy rate, as depicted in Figure 2 (right side), revealing that the cleaned dataset exhibited a 0.0% redundancy rate. This outcome indicated that the data was now prepared for rigorous analysis. The resultant clean dataset, complete with column names, was then exported, encompassing 149,043 rows, making it ready for further analysis.

TABLE II. Attributes of the UGRansome Dataset

| Attribute | Meaning | Type | Example |
|-----------|---------|------|---------|
| Time | Timestamp of network attacks | Numeric | 50s |
| Protocol | Network protocol | Categorical | TCP |
| Flag | Connection status | Categorical | ACK |
| Family | Ransomware family | Categorical | WannaCry |
| Clusters | Malware groups | Numeric | 1-12 |
| SeedAddress | Ransomware links | Categorical | 18y345 |
| ExpAddress | Ransomware links | Categorical | y7635d |
| BTC | Ransomware Bitcoin transactions | Numeric | 90.0 |
| USD | Ransomware USD transactions | Numeric | 32,465 |
| Netflow Bytes | Bytes transferred in network flow | Numeric | 45,389 |
| IP Address | IP addresses | Categorical | Class A |
| Threats | Malware | Categorical | Blacklist |
| Port | Network port number | Numeric | 5062 |
| Prediction | Outcomes of predictive models | Categorical | Anomaly (A) |

## 3. Designing the Ransomware Feature Selection Algorithm (RFSA)

Designing a novel feature selection algorithm for classifying ransomware transactions requires careful consideration of various factors and approaches [27]. This section outlines the proposed RFSA.

### Algorithm: RFSA

Objective: Select a subset of relevant features from a set of candidate attributes for classifying ransomware transactions from the UGRansome dataset.

Input: UGRansome

- $X$: The feature matrix, where each row represents a transaction, and each column represents a ransomware feature.

- $y$: The target labels, indicate whether each transaction is related to Anomaly (A), Signature (S), and Synthetic Signature (SS) (Table II).

- $k$: The desired number of selected features.

Output: A subset of the most relevant $k$ features.

### Algorithm

Feature Ranking: Calculate a ranking score for each feature based on its relevance to the classification task [27]. For each feature $i$:

$$\text{Score}(i) = \text{Gini Impurity}(X[:, i], y) \quad (1)$$

Select Top $k$ Features: Sort the features based on their ranking scores in descending order and select the top $k$ features [27]. Let $S$ be the set of selected features by the RFSA, and $S*$ be the optimal set of features that maximizes classification performance.

Relevance Ranking: RFSA calculates the relevance score for each feature based on a suitable relevance measure. By design, the higher the score, the more relevant the feature is to the extraction task [27]. To prove the algorithm's optimality, we need to show that $S$ is as close as possible to $S*$. The RFSA has been presented in Algorithm 1. The algorithm's optimality is based on its design, which prioritizes the selection of highly relevant features. The selected features $S$ are chosen to maximize the relevance score.

$$\text{Score}(i) \geq \text{Score}(j), \quad \forall i \in S, j \quad (2)$$

### A. Relevance Measure and Score Calculation

In Step 2 of Algorithm 1, we calculate the relevance score (Score$(i)$) for each feature $i$ using a suitable relevance measure [27]. A common relevance measure is Mutual Information [28], which quantifies the dependency between the feature and the target variable. The formula for Mutual Information is [29]:

$$\text{MI}(X_i, Y) = \sum_{x \in X_i} \sum_{y \in Y} p(x_i, y) \log \frac{p(x_i, y)}{p(x_i)p(y)} \quad (3)$$

## Algorithm 1 RFSA

**Require:** Feature matrix $X$, target labels $y$, desired number of selected features $k$
**Ensure:** Subset of top $k$ relevant features
1: **for** each feature $i$ in $X$ **do**
2:    Calculate Score($i$) using MI & GI
3: **end for**
4: Sort features in descending order based on Score($i$)
5: Select the top $k$ features as $f_k$
6: **return** $f_k$

where: - $X_i$ is the feature $i$ - $Y$ is the target variable - $p(x_i, y)$ is the joint probability distribution of $X_i$ and $Y$ - $p(x_i)$ and $p(y)$ are the marginal probability distributions of $X_i$ and $Y$, respectively. The MI score measures the amount of information shared between the feature and the target variable [28]. Higher scores indicate stronger dependencies [30]. The RFSA also used the Gini Impurity to measure the degree of disorder in the UGRansome dataset as follows:

$$GI(D) = 1 - \sum_{i=1}^{C} (p_i)^2 \qquad (4)$$

where: - $D$ represents the dataset. - $C$ is the number of classes in the dataset. - $p_i$ is the probability of an element in the dataset belonging to class $i$ [29]. The Gini Impurity Decrease quantifies the reduction in impurity achieved by splitting a dataset based on a particular feature and is calculated as follows:

$$GI_{\text{decrease}}(D, F) = GI(D) - \sum_{v \in \text{values}(F)} |D| \cdot GI(D_v) \qquad (5)$$

where: - $F$ is the feature being considered for the split. - $D_v$ represents the subset of data where feature $F$ takes the value $v$. To compute the importance of a feature, we consider its contribution to reducing Gini Impurity across multiple decision tree nodes. The feature importance score is calculated as follows:

$$FI(F) = \sum_{T_{t=1}} GI_{\text{decrease}}(D_t, F) \qquad (6)$$

where: - $FI(F)$ is the feature importance score for feature $F$. - $T$ represents the total number of decision tree nodes. - $D_t$ is the dataset at node $t$. - The denominator sums the Gini Impurity Decreases for feature $F$ across all nodes and features. The study used the Pearson Correlation Coefficient to measure the linear relationship between two variables and is calculated in Equation 7 [29]. where: - $\rho(X, Y)$ represents the Pearson Correlation Coefficient between variables $X$ and $Y$. - cov($X$, $Y$) is the covariance between $X$ and $Y$. - $\sigma x$ and $\sigma y$ are the standard deviations of $X$ and $Y$, respectively. The Correlation Matrix contains the pairwise correlations between different variables and is represented in Equation 8.

$$\rho(X, Y) = \frac{\text{cov}(X, Y)}{\sigma_X \cdot \sigma_Y} \qquad (7)$$

$$\text{Corr}(X, Y) =: \qquad (8)$$

$$\begin{matrix} 1 & \rho(X_1, Y_1) & \rho(X_1, Y_2) & ... & \rho(X_1, Y_n) \\ \rho(X_2, Y_1) & 1 & \rho(X_2, Y_2) & ... & \rho(X_2, Y_n) \\ - & - & - & \cdots & - \\ \rho(X_n, Y_1) & \rho(X_n, Y_2) & \rho(X_n, Y_3) & ... & 1 \end{matrix} \qquad (9)$$

where: - Corr($X$, $Y$) is the Correlation Matrix. - $\rho(X_i, Y_j)$ represents the Pearson Correlation Coefficient between variables $X_i$ and $Y_j$.

### B. Evaluation

Four evaluation metrics have been used to evaluate the performance of the RFSA (Equation 10).

$$\text{Accuracy} = \frac{\text{Number of Correct Predictions}}{\text{Total Number of Predictions}}$$

$$\text{Precision} =$$
$$\frac{\text{True Positives}}{\text{True Positives + False Positives}} \qquad (10)$$

$$\text{F1-Score} = \frac{2 \times \text{Precision} \times \text{Recall}}{\text{Precision + Recall}}$$

Accuracy measures the proportion of correctly classified instances to the total number of instances in the dataset [7]. It provides an overall view of how well the algorithm performs in terms of correct classifications. A higher accuracy indicates better performance. Precision measures the ratio of true positive predictions to the total number of positive predictions (both true positives and false positives) [6]. It evaluates the algorithm's ability to make accurate positive predictions.

A high precision indicates that the algorithm has fewer false positive errors. Recall, also known as sensitivity or true positive rate, measures the ratio of true positive predictions to the total number of actual positives (true positives and false negatives) [23]. It assesses the algorithm's ability to identify all positive instances correctly.

A high recall indicates that the algorithm can detect most of the positive cases. The F1-Score is the harmonic mean of precision and recall [26]. It provides a balanced evaluation of an algorithm's performance by considering both false positives and false negatives. It is particularly useful when dealing with imbalanced datasets. A higher

F1-Score indicates a better trade-off between precision and recall.

## 4. Results

Additional feature transformation techniques were subsequently employed on the initial dataset to facilitate the extraction and conversion of existing features into more actionable and informative variables. These transformed variables will be subjected to subsequent analysis and visualization. The forthcoming section provides a comprehensive discussion of the feature transformation techniques that were employed.

### A. Data Pre-Processing

Upon examination of the dataset insights provided by a Python DataPrep library [31], it became evident that three of the numerical features (namely, BTC, USD, and Netflow Bytes) exhibited significant skewness in their distributions (Figure 3). Consequently, a series of mathematical transformations [32] were implemented on these features to mitigate their skewed distributions, ultimately seeking to achieve either a normal distribution or a less-skewed distribution (Figure 5).

The logarithm [32] of each value of the feature is used in an attempt to normalize its distribution (un-skew it) (Figure 3 and Figure 5). This is one of the simpler mathematical transformation techniques applied and is especially useful in correcting features that are originally skewed to the right [32]. It assisted in centering the distribution of Netflow Bytes, which was originally skewed right ($\gamma_1 = 1.5737$). The value of 1 is added to each log to prevent zeros from occurring, as $\log(1)$ is equal to 0.

The final value used for analysis corresponds to the square root of each feature's values (Figure 3). This transformation is employed to normalize positively skewed distributions, particularly those skewed to the right [32]. This transformation was favored over the logarithmic approach for the USD feature due to its more pronounced centering effect (Figure 5). It is noteworthy that the initial distribution of the USD feature exhibited a right skewness ($\gamma_1 = 3.2318$).

The Yeo-Johnson transformation, which generalizes the Box-Cox transformation [32], is a mathematical technique that employs various power transformations (including logarithmic and inverse transformations) to modify a feature's data, aiming to make its distribution more normalized (Figure 3). Specifically, the Yeo-Johnson transformation adjusts low-variance data upward and high-variance data downward, while also accommodating negative values (Figure 3 and Figure 5). Figure 4 presents a histogram of the time feature along with various descriptive characteristics. The histogram reveals the following insights:

- Timestamp exhibits a slight right skewness (positively skewed), indicated by the mean being higher than the median.

- Approximately 68% of network attacks occur within the time range of 16.58 to 48.35, which corresponds to one standard deviation (SD) from the mean (mean± 1SD).

- The average timestamp of network attacks is 32.47 (mean).

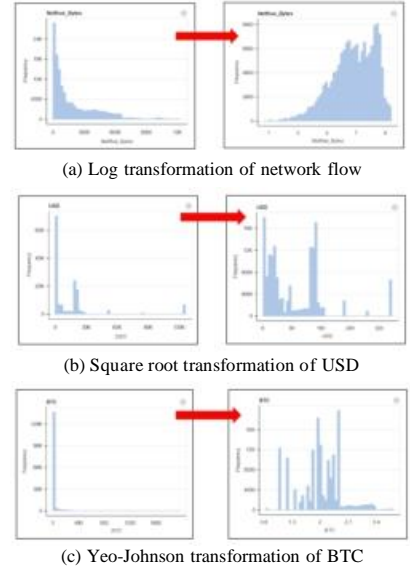

(a) Log transformation of network flow

(b) Square root transformation of USD

(c) Yeo-Johnson transformation of BTC

Figure 3. Numerical data transformation

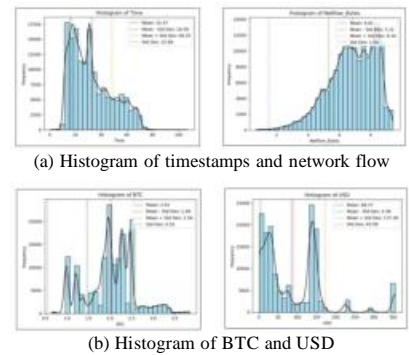

(a) Histogram of timestamps and network flow

(b) Histogram of BTC and USD

Figure 4. Histogram of transformed numerical attributes

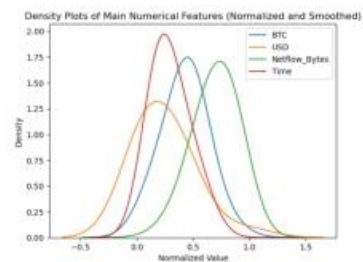

Figure 5. Normalized and smoothed numerical features

Figure 4 depicts a histogram of the BTC feature along with various descriptive characteristics. The histogram yields the following observations:

- BTC exhibits a very slight left skewness (negatively skewed), as evidenced by the mean being lower than the median.

- Approximately 68% of attacks involve BTC transactions within the range of 1.46 to 2.56, which corresponds to one standard deviation from the mean (mean ± 1SD).

- The average number of BTC transactions per attack is 2.01 (mean).

- There are potential outliers in the range of 0.5 to 1.0 BTC transactions, represented by bins with lower counts and distinct separation from the main distribution.

Figure 4 displays a histogram of the USD feature alongside various descriptive characteristics. The histogram reveals the following insights:

- USD exhibits a slight right skewness (positively skewed), as indicated by the mean being higher than the median.

- Approximately 68% of attacks resulted in financial damages ranging from 4.38 to 172.36 USD, which corresponds to one standard deviation from the mean (mean ± 1SD).

- The average financial damage per attack is 88.37 USD (mean).

- There are significant outliers in the range of 200 to 300 USD, represented by bins with lower counts and distinct separation from the main distribution.

Furthermore, categorical variables were proficiently converted into numerical equivalents, making them suitable for a wide range of modeling and analytical methodologies (Figure 6). This enriched dataset, now composed of numeric representations, becomes a valuable resource in the context of feature extraction.

### B. Ransomware Classification

Table III portrays the RFSA results. Figure 7 illustrates feature importance as determined by Gini Impurity. In the context of feature importance, Gini Impurity quantifies how well a feature separates classes or categories within the dataset. Features that lead to better separation and lower impurity are considered more important as they contribute more to the decision-making process in the extraction tasks. The fluctuation in the performance metrics based on the selected ransomware features provides valuable insights into how each feature impacts the extraction of ransomware transactions (Figure 8). The accuracy is slightly higher when the USD feature is selected compared to the BTC

(a) Original dataset: categorical vs. numerical features

(b) Encoded dataset: numerical features

Figure 6. The original and encoded dataset

feature. This suggests that using USD as a feature yields a more accurate model for the extraction of ransomware transactions (Figure 8). The precision is higher for BTC, indicating that when BTC is included as a feature, the model is better at correctly extracting positive cases of ransomware transactions. BTC also leads in the recall, meaning it captures more true positive cases, which is essential for identifying ransomware transactions (Figure 8).

Figure 7. Feature importance

The F1 score considers both precision and recall and shows a slight advantage for BTC. These three features, clusters, port, and address 1SYSTEMQ, have relatively close scores in all metrics (Figure 8). This suggests that they contribute similarly to the extraction task, and the choice between them may depend on other considerations like computational efficiency or domain knowledge. The MI score decreases as we move down the selected features. This indicates that USD provides the most information gain, followed by BTC, clusters, port, and address 1SYSTEMQ (Figure 8 and Figure 7). Features with higher MI scores are generally more informative for extraction, as they are more relevant to distinguishing between ransomware and non-ransomware transactions.

TABLE III. Feature Selection and Evaluation Metrics

| Selected Features | Number of Features | Target | MI Score | Accuracy (%) | Precision (%) | Recall (%) |
|---|---|---|---|---|---|---|
| USD | 12,000 | Anomaly | 95.6 | 93.2 | 89.5 | 92.8 |
| BTC | 11,800 | Signature | 92.4 | 92.7 | 91.0 | 93.5 |
| Clusters | 11,500 | Synthetic Signature | 89.3 | 91.5 | 90.2 | 91.8 |
| Port | 11,200 | Anomaly | 87.2 | 91.1 | 89.8 | 92.3 |
| address  1SYSTEMQ | 11,050 | Signature | 85.0 | 90.3 | 88.7 | 92.1 |
| Flag  APSF | 11,030 | Synthetic Signature | 82.9 | 90.1 | 88.5 | 92.0 |
| address  1GZkujBR | 11,020 | Anomaly | 80.7 | 89.9 | 88.2 | 91.9 |
| Flag  AF | 11,010 | Signature | 78.5 | 89.6 | 87.9 | 91.7 |
| Protocol  TCP | 11,005 | Synthetic Signature | 76.3 | 89.4 | 87.5 | 91.6 |
| DoS | 11,001 | Synthetic Signature | 69.7 | 88.7 | 86.4 | 91.3 |
| UDP | 11,000 | Anomaly | 67.5 | 88.4 | 86.1 | 91.1 |
| ICMP | 10,990 | Synthetic Signature | 63.1 | 88.0 | 85.4 | 90.9 |
| address  18e372GN | 10,985 | Anomaly | 60.9 | 87.7 | 85.0 | 90.8 |
| address  1NKi9AK5 | 10,980 | Signature | 58.7 | 87.5 | 84.6 | 90.6 |
| Globe | 10,975 | Synthetic Signature | 56.5 | 87.2 | 84.3 | 90.5 |
| address  17dcMo4V | 10,970 | Anomaly | 54.3 | 87.0 | 83.9 | 90.4 |
| Scan | 10,960 | Synthetic Signature | 49.9 | 86.5 | 83.2 | 90.1 |
| Spam | 10,955 | Anomaly | 47.7 | 86.2 | 82.8 | 90.0 |
| address  1BonuSr7 | 10,950 | Signature | 45.5 | 86.0 | 82.4 | 89.8 |
| SamSam | 10,945 | Synthetic Signature | 43.3 | 85.7 | 82.1 | 89.7 |
| SSH | 10,940 | Anomaly | 41.1 | 85.5 | 81.7 | 89.5 |
| Blacklist | 10,925 | Anomaly | 34.5 | 84.7 | 80.6 | 89.1 |
| Botnet | 10,920 | Signature | 32.3 | 84.5 | 80.2 | 88.9 |
| Bonet | 10,915 | Synthetic Signature | 30.1 | 84.2 | 79.9 | 88.8 |
| APT | 10,910 | Anomaly | 27.9 | 84.0 | 79.5 | 88.6 |
| Locky | 10,905 | Signature | 25.7 | 83.7 | 79.1 | 88.5 |
| NerisBonet | 10,900 | Synthetic Signature | 23.5 | 83.5 | 78.8 | 88.3 |
| TowerWeb | 10,895 | Anomaly | 21.3 | 83.2 | 78.4 | 88.2 |
| address  1LC7xTpP | 10,890 | Signature | 19.1 | 83.0 | 78.0 | 88.0 |
| EDA2 | 10,885 | Synthetic Signature | 16.9 | 82.7 | 77.7 | 87.9 |
| Flyper | 10,880 | Anomaly | 14.7 | 82.5 | 77.3 | 87.7 |
| Razy | 10,875 | Signature | 12.5 | 82.2 | 76.9 | 87.6 |
| Cryptohitman | 10,870 | Synthetic Signature | 10.3 | 82.0 | 76.6 | 87.4 |
| JigSaw | 10,865 | Anomaly | 8.1 | 81.7 | 76.2 | 87.3 |
| address  1AEoiHYZ | 10,860 | Signature | 5.9 | 81.5 | 75.8 | 87.1 |
| WannaCry | 10,855 | Synthetic Signature | 3.7 | 81.2 | 75.5 | 87.0 |
| CryptXXX | 10,850 | Anomaly | 1.5 | 81.0 | 75.1 | 86.8 |
| DMALocker | 10,845 | Signature | 0.3 | 80.7 | 74.7 | 86.7 |
| NoobCrypt | 10,840 | Synthetic Signature | 0.1 | 80.5 | 74.4 | 86.5 |
| address  1KZkcvx4 | 10,835 | Anomaly | 0.0 | 80.2 | 74.0 | 86.4 |
| CryptoLocker | 10,830 | Signature | 0.0 | 80.0 | 73.6 | 86.2 |
| Globev3 | 10,825 | Synthetic Signature | 0.0 | 79.7 | 73.3 | 86.1 |

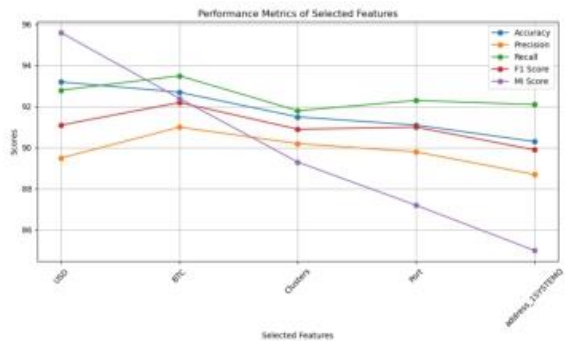

Figure 8. Performance metrics of selected features

## C. Implication

The choice of features significantly impacts the performance of a ransomware feature extraction model. The USD and BTC appear to be the most influential features, as they consistently perform well across all metrics. While BTC excels in precision and recall, USD achieves a slightly higher accuracy. The choice between these two features may depend on the specific objectives and trade-offs in a real-world application. It is essential to consider both the MI score and individual metric performance when selecting features.

Features with higher MI scores are likely to have a more substantial impact on the model's performance. In summary, the fluctuation in performance metrics provides guidance on feature selection for ransomware detection. The choice of features should align with the specific goals of the extraction task, considering factors such as accuracy, precision, recall, and the MI score. A combination of features may also be beneficial in achieving a balanced trade-off between different aspects of model performance.

The categorical data of extracted features exhibits an evident class imbalance, as depicted in Figure 9. This graph visually presents the distribution of various ransomware types, revealing discrepancies among them. Specifically, it shows that the Locky ransomware class is more prevalent than the Globev3 ransomware class. Consequently, even though there are 17 unique classes, the dataset demonstrates a substantial imbalance, with a small number of classes accounting for the majority of the data. However, it is important to note that the data's overall shape remains consistent with the original dataset. The reduction in certain instances is primarily due to the removal of outliers and duplicates, which has helped slightly balance the dataset. This process is depicted in Figure 10.

The stacked bar chart presented in Figure 11 provides a comprehensive view of the prediction distribution across different threat or malware categories. Among the nine malware categories, SSH stands out with the highest bar, primarily due to its substantial count within the dataset. However, it is important to emphasize that this high count does not necessarily convey any predictive information (see Figure 11). The predictive variable assigned to each entry categorizes it as either a well-known threat, denoted as Signature (S), or an unknown and potentially zero-day threat or anomaly, indicated as Anomaly (A) or Synthetic Signature (SS).

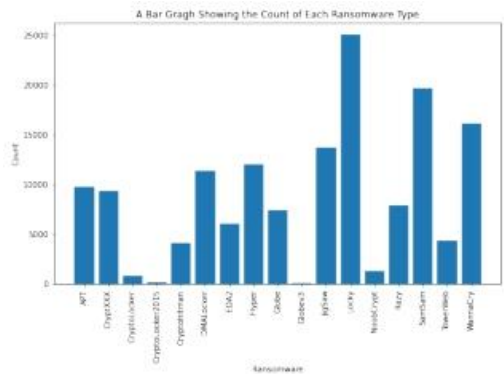

Figure 9. Extracted ransomware families

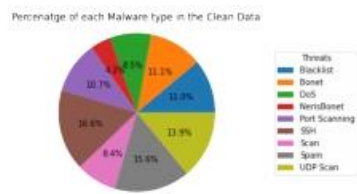

Figure 10. Extracted malware

Examining the graph, we observe that categories like Blacklist, Port Scanning, and Spam are predominantly associated with well-known threats, with relatively few anomalies and synthetic signatures. This suggests that the occurrence of one-day threats or anomalies in these categories is less likely. In contrast, the other malware types, when considering the count of abnormal attacks, signal a higher likelihood of zero-day threat scenarios emerging from these categories.

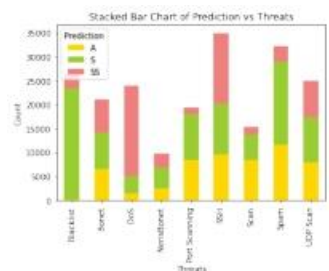

Figure 11. Threat prediction

In Figure 12, we can observe the average time it takes for a particular malware type to infiltrate an organization's network, measured in seconds. This data provides valuable insights into the varying degrees of efficiency exhibited by different malware types when it comes to breaching network defenses. The graph reveals that all nine categories of malware exhibit similar average infiltration times. However, an intriguing pattern emerges when we consider the threats previously identified as having a high percentage of safe signatures, namely Blacklist, Port Scanning, and Spam.

These threats appear to be the quickest at breaching an organization's network, contrasting with the other malware types categorized as unknown threats, which, on average, require more time to infiltrate the network. Among these, the Bonet malware type stands out as having the longest average infiltration time. Furthermore, the malware types can be further grouped into different ransomware types, as illustrated in Figure 13, a stacked bar chart displaying the 17 ransomware types and their respective malware counts.

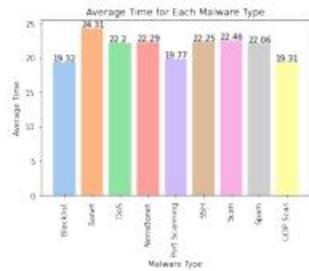

Figure 12. Average timestamp of each ransomware

Locky ransomware, known for encrypting files and demanding a BTC ransom for decryption, has the highest overall count. Locky ransomware is primarily composed of SSH, Scan, and UDP Scan malware, although it exhibits associations with every malware type. This finding has significant implications for assessing the likelihood of a successful network attack targeting an organization.

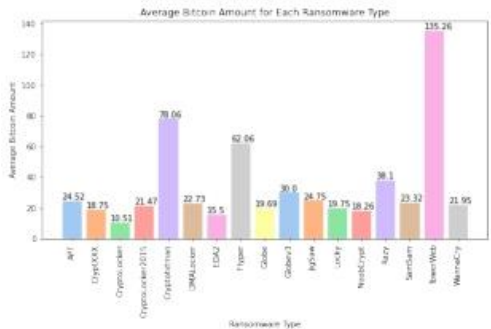

Figure 13. Financial impact of malware

Lastly, it is worth highlighting that not all malware types are intricately linked to specific ransomware categories. For instance, CryptoLocker2015 is exclusively associated with one type of malware, Blacklist, which potentially limits its effectiveness in executing a successful attack on an organization's network. Moreover, Blacklist is often predicted to be recognized as a signature attack, further hindering its infiltration. Another intriguing aspect of the dataset involves examining the average ransom prices associated with each ransomware type. Towerweb emerges as the ransomware demanding the highest fee in terms of BTC, amounting to 135.26, in stark contrast to CryptoLocker, which commands the lowest fee at 10.51 (Figure 14). This insight sheds light on the considerable variation in ransom demands across different ransomware types.

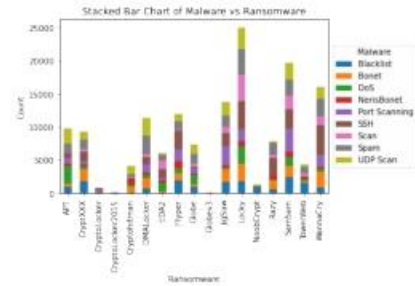

Figure 14. Selected malware and ransomware

The correlation matrix of the extracted features is visually represented in Figure 15. It reveals a significant correlation coefficient of 0.26 between the ransomware cluster and predicted BTC transactions. This finding underscores a robust association between specific ransomware attack types and distinctive patterns within cryptocurrency transactions.

For instance, if we consider a scenario in which the Locky ransomware cluster consistently demands BTC payments as ransom. The pronounced correlation observed suggests that analyzing BTC transaction patterns can serve as a practical approach to identifying and forecasting Locky ransomware attacks.

Security systems and machine learning models can harness this correlation to bolster their detection and response mechanisms, ultimately enhancing their capacity to thwart ransomware incidents and fortify defenses against cyber threats effectively.

Moreover, Figure 16 provides valuable insights into the intricate relationship between ransomware timestamp and the variables USD, BTC, and Netflow Bytes. Essentially, it addresses the question of how the duration of a ransomware attack impacts financial gains and the volume of Netflow Bytes. The visualizations clarify that, generally, a more extended duration corresponds to higher financial gains, but this correlation does not guarantee substantial gains, with the trend typically commencing around a time value of 2.5, except for a few outliers.

The same pattern emerges concerning Netflow Bytes, emphasizing not only the connection between timestamp and USD, BTC, and Netflow Bytes but also the pivotal role of increased Netflow Bytes in achieving financial gains. The graphs reveal that the significant gains in currency and Netflow Bytes predominantly occur within the time interval of 2.5 to 4.5. This observation leads us to predict that during a ransomware attack, these time intervals are critical junctures for assessing potential financial gains and gauging the flow of Netflow Bytes (Figure 17).

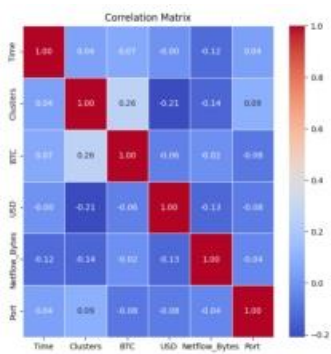

Figure 15. Correlation matrix of extracted features

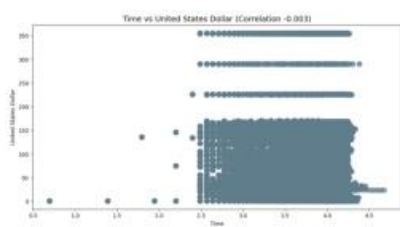

(a) Timestamp and USD correlation

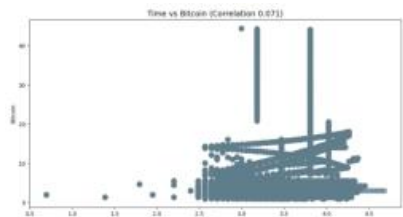

(b) Timestamp and BTC correlation

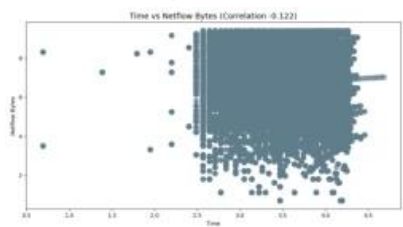

(c) Timestamp and Network Flow correlation

Figure 16. Numerical feature correlation

Figure 18 provides a comprehensive analysis of the financial gains in USD associated with ransomware attacks based on the originating port or utilized protocol. It also offers insights into the financial gains influenced by the specific ransomware family or malware threat in conjunction with the port or protocol. These visualizations offer the means to predict the potential success of an attack by considering factors such as the ransomware family or threat type alongside the port or protocol used. For instance, it is notable that port 5066 yields the highest financial gains in

USD at an earlier time point, whereas port 5068 leads to the highest gains at a later time point. Among the pairings of protocol and threat, the TCP protocol paired with the NerisBonet threat stands out as the most successful, while the combination of Port 5068 and the Spam threat emerges as highly effective.

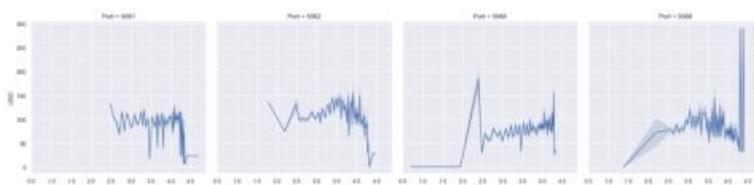

Figure 17. Attack timestamp prediction

In terms of ransomware and protocol pairings, the TCP protocol combined with the NoobCrypt family is successful, as is the combination of port 5068 with the NoobCrypt family. Consequently, these findings suggest that the most successful attacks tend to originate from port 5068 or employ the TCP protocol, with the NoobCrypt family exhibiting proficiency in both scenarios.

## 5. Discussion

In conclusion, the financial aspects of ransomware attacks revealed a lack of a clear-cut relationship between ransomware types and the associated ransom amounts in BTC. This observation underscores the variability in the ransom demands across different ransomware families, suggesting that there is no fixed or predetermined amount for a particular type of cyberattack. The ransomware landscape remains dynamic and adaptable, with threat actors continuously adjusting their ransom demands. Furthermore, the analysis delved into the correlation matrix of extracted features, revealing a noteworthy correlation of 0.26 between ransomware clusters and predicted BTC transactions. This correlation signifies a robust association between specific ransomware attack types and distinctive patterns in cryptocurrency transactions. For instance, the high correlation suggests that monitoring BTC transaction patterns can serve as a practical means of identifying and predicting ransomware attacks, such as the Locky ransomware. Leveraging this correlation can enhance the effectiveness of security systems and machine learning models, leading to improved detection and response mechanisms, and ultimately bolstering cybersecurity defenses against ransomware threats.

Additionally, the examination of temporal aspects, particularly the relationship between attack duration and gains in currency (USD and BTC), shed light on critical time intervals during ransomware attacks. The analysis indicated that the most significant gains in currency typically occurred between specific time points, highlighting the importance of monitoring and responding to threats during these critical phases. In summary, this comprehensive analysis of ransomware-related data provides valuable insights into the dynamic and evolving nature of cyber threats. It emphasizes the need for adaptable cybersecurity strategies and proactive measures that leverage data-driven approaches to mitigate the risks posed by ransomware attacks. Figure 19 illustrates the relationship between Gini Impurity and MI scores for various ransomware classes categorized into Signature (S), Synthetic Signature (SS), and Anomaly (A). Each ransomware class exhibits distinct patterns, with TowerWeb displaying higher MI scores, indicating more predictable web-based transaction behaviors. Conversely, NoobCrypt demonstrates greater variability in both criteria. These dynamics underscore the need for adaptive detection methods to account for evolving web and cryptographic ransomware behaviors. Insights gleaned from this graph suggest that feature selection based on Gini Impurity and MI can effectively discriminate between ransomware classes, which has significant implications for improving ransomware detection and classification systems. Understanding these dynamics can contribute to the development of more accurate and adaptive machine learning models, enhancing cybersecurity efforts against web-based cryptographic threats.

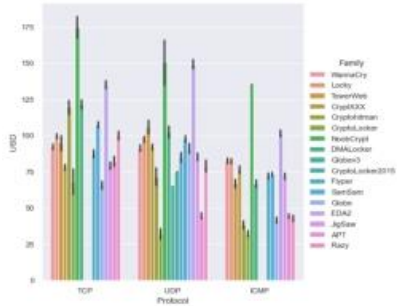

(a) USD and ransomware protocol

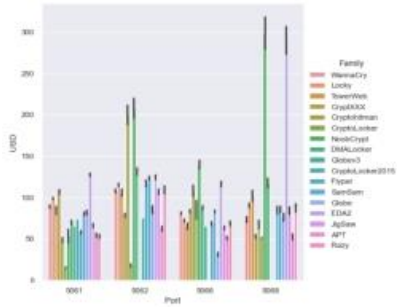

(b) USD and ransomware ports

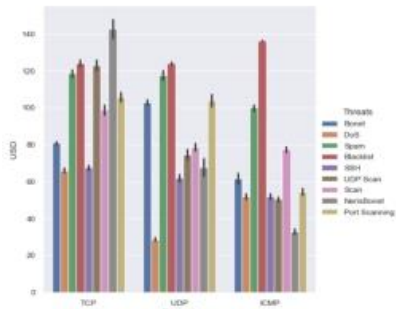

(c) USD and malware protocols

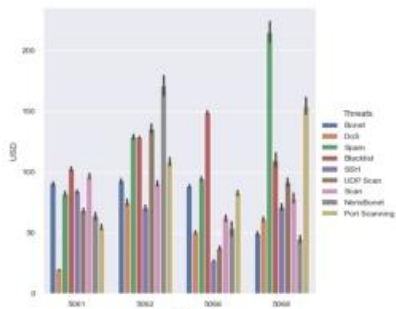

(d) USD and malware ports

Figure 18. Malware extracted

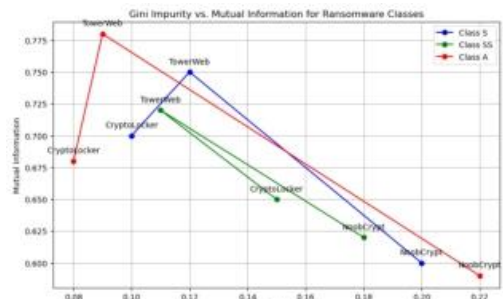

Figure 19. Gini Impurity and MI scores

## 6.  Conclusion

In conclusion, this study provides a multifaceted analysis of ransomware-related data, offering insights that underscore the complexity and evolving nature of cybersecurity threats. Our exploration of financial aspects revealed the absence of a fixed ransom amount associated with specific ransomware types, highlighting the adaptability of threat actors. Moreover, the correlation analysis unveiled a strong link between ransomware clusters and cryptocurrency transaction patterns, enhancing the potential for predictive and preventive cybersecurity measures. The temporal analysis emphasized critical time intervals during ransomware attacks, guiding the development of timely response strategies. Collectively, these findings emphasize the importance of data-driven, adaptive cybersecurity approaches to effectively address the ever-changing landscape of ransomware threats, safeguarding organizations and individuals against potential cyberattacks.

## References

[1]  C. Leuprecht, C. Jenkins, and R. Hamilton, "Virtual money laundering: policy implications of the proliferation in the illicit use of cryptocurrency," *Journal of Financial Crime*, vol. 30, no. 4, pp. 1036–1054, 2023.

[2]  D. Chaudhari, R. Agarwal, and S. K. Shukla, "Towards malicious address identification in bitcoin," in *2021 IEEE International Conference on Blockchain (Blockchain)*.   IEEE, 2021, pp. 425–432.

[3]  A. Alqahtani and F. T. Sheldon, "A survey of crypto ransomware attack detection methodologies: an evolving outlook," *Sensors*, vol. 22, no. 5, p. 1837, 2022.

[4] A. Zimba, Z. Wang, and H. Chen, "Multi-stage crypto ransomware attacks: A new emerging cyber threat to critical infrastructure and industrial control systems," *Ict Express*, vol. 4, no. 1, pp. 14–18, 2018.

[5] M. Paquet-Clouston, B. Haslhofer, and B. Dupont, "Ransomware payments in the bitcoin ecosystem," *Journal of Cybersecurity*, vol. 5, no. 1, p. tyz003, 2019.

[6] M. Nkongolo and M. Tokmak, "Zero-day threats detection for critical infrastructures," in *South African Institute of Computer Scientists and Information Technologists*, A. Gerber and M. Coetzee, Eds. Cham: Springer Nature Switzerland, 2023, pp. 32–47.

[7] M. Nkongolo, J. P. Van Deventer, and S. M. Kasongo, "Ugransome1819: A novel dataset for anomaly detection and zero-day threats," *Information*, vol. 12, no. 10, p. 405, 2021.

[8] N. Ganesh, R. Shankar, R.ˇCep, S. Chakraborty, and K. Kalita, "Efficient feature selection using weighted superposition attraction optimization algorithm," *Applied Sciences*, vol. 13, no. 5, p. 3223, 2023.

[9] M. Lichtenstein and Z. Rucks-Ahidiana, "Contextual text coding: A mixed-methods approach for large-scale textual data," *Sociological Methods & Research*, vol. 52, no. 2, pp. 606–641, 2023.

[10] J. Han, J. Pei, and H. Tong, *Data mining: concepts and techniques*. Morgan kaufmann, 2022.

[11] M. Nkongolo, J. P. Van Deventer, S. M. Kasongo, S. R. Zahra, and J. Kipongo, "A cloud based optimization method for zero-day threats detection using genetic algorithm and ensemble learning," *Electronics*, vol. 11, no. 11, p. 1749, 2022.

[12] W. Hardy, L. Chen, S. Hou, Y. Ye, and X. Li, "Dl4md: A deep learning framework for intelligent malware detection," in *Proceedings of the International Conference on Data Science (ICDATA)*. The Steering Committee of The World Congress in Computer Science, Computer ..., 2016, p. 61.

[13] P. Panda, O. K. CU, S. Marappan, S. Ma, and D. Veesani Nandi, "Transfer learning for image-based malware detection for iot," *Sensors*, vol. 23, no. 6, p. 3253, 2023.

[14] Y. Fang, C. Huang, L. Liu, and M. Xue, "Research on malicious javascript detection technology based on lstm," *IEEE Access*, vol. 6, pp. 59118–59125, 2018.

[15] C. Roberts and M. Nair, "Arbitrary Discrete Sequence Anomaly Detection with Zero Boundary LSTM," *arXiv e-prints*, p. arXiv:1803.02395, Mar. 2018.

[16] T. Wang, W. W. Y. Ng, W. Li, S. Kwong, and J. Li, "Broad autoencoder features learning for pattern classification problems," in *2019 IEEE 18th International Conference on Cognitive Informatics Cognitive Computing (ICCI*CC)*, 2019, pp. 130–135.

[17] S. Chatterjee, D. Dey, and S. Munshi, "Morphological, texture and auto-encoder based feature extraction techniques for skin disease classification," in *2019 IEEE 16th India Council International Conference (INDICON)*, 2019, pp. 1–4.

[18] X. Kong, R. Lin, and H. Zou, "Feature extraction of load curve based on autoencoder network," in *2020 IEEE 20th International Conference on Communication Technology (ICCT)*, 2020, pp. 1452–1456.

[19] Y. Wang, H. Yang, X. Yuan, Y. A. Shardt, C. Yang, and W. Gui, "Deep learning for fault-relevant feature extraction and fault classification with stacked supervised auto-encoder," *Journal of Process Control*, vol. 92, pp. 79–89, 2020. [Online]. Available: https://www.sciencedirect.com/science/article/pii/S0959152420302225

[20] J. Kim, H. Lee, J. W. Jeon, J. M. Kim, H. U. Lee, and S. Kim, "Stacked auto-encoder based cnc tool diagnosis using discrete wavelet transform feature extraction," *Processes*, vol. 8, no. 4, 2020. [Online]. Available: https://www.mdpi.com/2227-9717/8/4/456

[21] M. Tokmak, "Deep forest approach for zero-day attacks detection," *Innovations and Technologies in Engineering.*, no. ISBN: 978-625-6382-83-1, pp. 45–56, 2022.

[22] A. Jyothish, A. Mathew, and P. Vinod, "Effectiveness of machine learning based android malware detectors against adversarial attacks," *Cluster Computing*, pp. 1–21, 2023.

[23] D. Shankar, G. V. S. George, J. N. J. N. S. S, and P. S. Madhuri, "Deep analysis of risks and recent trends towards network intrusion detection system," *International Journal of Advanced Computer Science and Applications*, vol. 14, no. 1, 2023.

[24] F. Suthar, N. Patel, and S. Khanna, "A signature-based botnet (emotet) detection mechanism," *Int. J. Eng. Trends Technol*, vol. 70, no. 5, pp. 185–193, 2022.

[25] A. Singh, Z. Mushtaq, H. A. Abosaq, S. N. F. Mursal, M. Irfan, and G. Nowakowski, "Enhancing ransomware attack detection using transfer learning and deep learning ensemble models on cloud-encrypted data," *Electronics*, vol. 12, no. 18, p. 3899, 2023.

[26] A. Rege and R. Bleiman, "A free and community-driven critical infrastructure ransomware dataset," in *Proceedings of the International Conference on Cybersecurity, Situational Awareness and Social Media*, C. Onwubiko, P. Rosati, A. Rege, A. Erola, X. Bellekens, H. Hindy, and M. G. Jaatun, Eds. Singapore: Springer Nature Singapore, 2023, pp. 25–37.

[27] A. Thakkar and R. Lohiya, "Fusion of statistical importance for feature selection in deep neural network-based intrusion detection system," *Information Fusion*, vol. 90, pp. 353–363, 2023.

[28] P. Teisseyre and J. Lee, "Multilabel all-relevant feature selection using lower bounds of conditional mutual information," *Expert Systems with Applications*, vol. 216, p. 119436, 2023.

[29] Y. Yuan, L. Wu, and X. Zhang, "Gini-impurity index analysis," *IEEE Transactions on Information Forensics and Security*, vol. 16, pp. 3154–3169, 2021.

[30] M. Zhou, K. Yan, J. Huang, Z. Yang, X. Fu, and F. Zhao, "Mutual information-driven pan-sharpening," in *Proceedings of the IEEE/CVF Conference on Computer Vision and Pattern Recognition*, 2022, pp. 1798–1808.

[31] J. Peng, W. Wu, B. Lockhart, S. Bian, J. N. Yan, L. Xu, Z. Chi, J. M. Rzeszotarski, and J. Wang, "Dataprep. eda: Task-centric exploratory data analysis for statistical modeling in python," in *Proceedings of the 2021 International Conference on Management of Data*, 2021, pp. 2271–2280.

[32] J. Osborne, "Notes on the use of data transformations," *Practical assessment, research, and evaluation*, vol. 8, no. 1, p. 6, 2002.